\documentclass[%
 preprint,
 amssymb,
 aps,prd,
 floatfix,
 ]{revtex4-1}

\usepackage[centertags]{amsmath}
\usepackage{enumitem}

\newlist{STEP}{enumerate}{1}
\setlist[STEP]{label=\Roman*:}

\usepackage{dcolumn}% Align table columns on decimal point
\usepackage{bm}% bold math

\usepackage{amsmath}
\usepackage{graphicx}

\usepackage{hyperref}% add hypertext capabilities
\hypersetup{
	colorlinks=true,       % false: boxed links; true: colored links
	linkcolor=blue,          % color of internal links
	citecolor=blue,        % color of links to bibliography
	filecolor=blue,      % color of file links
	urlcolor=magenta           % color of external links
}

\usepackage{tikz}
\usetikzlibrary{positioning}
\usetikzlibrary{shapes,arrows,arrows,positioning,fit}
\tikzstyle{decision} = [diamond, draw, fill=blue!20, 
  text width=5.5em, text badly centered, inner sep=0pt]
  \tikzstyle{block} = [rectangle, draw, fill=green!20, 
  text width=30em, text centered, rounded corners, minimum height=4em]
  \tikzstyle{line} = [draw, -latex']
  \tikzstyle{cloud} = [draw, ellipse,fill=red!20,
  minimum height=4em]
  \tikzstyle{decisionn} = [diamond, draw, fill=red!20, 
  text width=4em, text badly centered, node distance=4cm, inner sep=0pt]
  \tikzstyle{blockk} = [rectangle, draw, fill=blue!20, 
  text width=8em, text centered, rounded corners, minimum height=2em]
  \tikzstyle{blockkk} = [rectangle, draw, fill=blue!20, 
  text width=8em, text centered, rounded corners, minimum height=2em]
  
\allowdisplaybreaks
\definecolor{darkblue}{rgb}{0.0, 0.0, 0.62}
\definecolor{deepmagenta}{rgb}{0.7, 0.01, 0.7}
\definecolor{darkred}{rgb}{0.55, 0.0, 0.0}
\graphicspath{{./graphs/}}

\usepackage{caption}
\usepackage{subcaption}
\makeatletter
\renewcommand\p@subfigure{}
\makeatother

\usepackage{cleveref}

\usepackage{subcaption}  % For subfigures/tables
\begin{document}
\title{Closed Timelike Curves from a Vacuum Traveling Wave}
%%%
\author{Semin Xavier}
\email{spssx3291@iacs.res.in}

\affiliation{%
 School of Physical Sciences, Indian Association for the Cultivation of Science, Kolkata-700032, India
}%

\date{\today}
\begin{abstract}
We construct an exact vacuum spacetime that develops closed timelike curves from regular, asymptotically flat initial data respecting the weak, dominant, and strong energy conditions. Einstein's equations fix the geometry through two functions, a harmonic transverse profile and a traveling-wave mode, whose evolution drives the closed curves of the time-machine core from spacelike to timelike. The chronology horizon admits a degenerate limit in which its generating closed null geodesic has vanishing boost and optical scalars, the conditions for a Killing spinor.

%We construct an exact vacuum spacetime that develops closed timelike curves from regular, asymptotically flat initial data satisfying the weak, dominant, and strong energy conditions. The geometry is governed by a harmonic transverse profile and a traveling-wave mode determined by Einstein’s equations, which drives a family of closed curves through a causal transition. The resulting chronology horizon possesses a degenerate limit in which the generating a closed null geodesic has vanishing boost and optical scalars and admits a Killing spinor.

\end{abstract}

%\pacs{}

\maketitle
\sloppy

\section{Introduction}
%%%%%%%%%%%%%%%%%%%%%%% Introduction %%%%%%%%%%%%%%%%%%%%%%%%
Unlike in Newtonian physics, the geometric structure of spacetime in general relativity(GR) can, in principle, permit violations of chronology and causality. Such violations, associated with the existence of closed timelike curves (CTCs), are commonly interpreted as \textit{time machines}. The existence of CTCs challenges the foundations of GR~\cite{2002_Visser}.

An extensive body of literature has examined the existence of spacetimes admitting CTCs. Early examples include Gödel's cosmological solution and the cylindrical spacetimes of Van Stockum and Tipler~\cite{1949-Godel-RevMPhy,1937-vanStockum-ProRSoc,1974-Tipler-PRD}. Subsequent studies have shown that chronology-violating structures can arise in a broad range of geometries, including the interior regions of Kerr and Kerr–Newman black holes~\cite{1973-Hawking_Ellis-Book}, Gott's moving cosmic-string configuration~\cite{1990-Gott-PRL}, and traversable wormhole spacetimes~\cite{1988-Morris_Thorne-PRL} (see Ref.~\cite{2002_Visser} for a more extensive list of such geometries). Although these examples possess distinct limitations, such as unrealistic global properties, the requirement of exotic matter that violates the classical energy conditions, or the presence of unstable horizons, the recurrence of chronology-violating behavior across diverse settings suggests that it is not merely an isolated feature of a particular solution~\cite{2007-Ori-PRD}. This observation has motivated considerable interest in the possibility of a general mechanism enforcing chronology protection, as well as in the question of whether time machines could arise in physically realizable situations.

A physically acceptable time-machine model was proposed by Ori~\cite{2005-Ori-PRL}, who constructed an asymptotically flat and topologically trivial spacetime in which CTCs develop from an initially causally well-behaved region while satisfying the weak, dominant, and strong energy conditions. This result demonstrated that the classical energy conditions alone do not necessarily prevent the formation of closed timelike curves. Nevertheless, the matter content supporting the geometry was not associated with any known fundamental field.
%, leaving open the question of whether a realistic dynamical mechanism can generate such spacetimes. 
Although the model exhibits a vacuum toroidal core in which closed timelike curves appear once the metric coefficient  $g_{zz}=f(x,y,z)-T$ changes sign, $T$ is a coordinate of a fixed geometry rather than a parameter that can be physically advanced, so the curves are a permanent feature of the solution rather than the product of a controllable process. The model therefore demonstrates the geometric existence of such a spacetime, but not a physical mechanism by which it could be dynamically generated or built, and its stability remains open. %It has nonetheless been suggested that, since the required initial data respect the energy conditions, they might in principle be established by a advanced civilization capable of manipulating spacetime on the relevant scales—or might conceivably arise from natural processes involving gravitating masses in rapid motion—although such prospects remain entirely speculative.

Whether such a spacetime can be generated at all is further constrained by the manner of generation itself, a connection made precise by Hawking's criterion of compact generation~\cite{1992-Hawking-PRD}. A chronology horizon, the boundary at which causality violation sets, is said to be compactly generated when its null generators, traced into the past, remain confined to a compact region of spacetime, winding repeatedly onto a closed, or almost closed, null geodesic. Such a horizon corresponds to a time machine that is genuinely \emph{built} within a finite region, the closed causal curves arising from a self-contained, bounded evolution rather than from structure imported from a singularity or from infinity. Hawking showed, however, that this very mode of generation is costly: when the generators are trapped and forced to reconverge, the Raychaudhuri equation renders their behaviour incompatible with the weak energy condition, so that a compactly generated time machine arising from a non-compact, asymptotically flat partial Cauchy surface must violate it~\cite{1992-Hawking-PRD}. The generation mechanism therefore controls the applicability of Hawking's criterion: a model whose causality violation develops through a compactly generated horizon necessarily requires exotic matter, whereas Ori's construction evades this conclusion precisely because its mechanism is \emph{not} compactly generated. There, the weak-energy-condition–respecting matter forces the geometry to develop a non-compactness
%—a singularity or an internal infinity—at late times, 
from which the horizon generators emanate, while only the initial data responsible for the onset of causality violation remain confined to a compact set. It is this distinction, relaxing the compact generation of the horizon while retaining a compact region of initial data,that allows the model to generate closed timelike curves without violating the energy conditions.

In this work, we construct a vacuum solution of the Einstein field equations and employ it as the core geometry of a new time-machine model. The solution is governed by two functions: a transverse harmonic function, which controls the local curvature, and a traveling-wave function, whose evolution fixes the causal character of the CTCs. We show that closed timelike curves arise dynamically as the metric component $g_{zz}$ changes sign, while the initial data, specified on a regular, asymptotically flat, and topologically trivial partial Cauchy surface, respect the weak, dominant, and strong energy conditions. We then analyze the causal geodesics that generate the chronology horizon and identify a degenerate tuning, so that the horizon generator admits a Killing spinor.

\section{Model and exact solution}
%%%%%%%%%%%%%% main content %%%%%%%%%%%%
We now proceed to describe the geometry of the time-machine core, starting from a vacuum solution,
{\small
\begin{equation}\label{ansatz}
    ds^2=-2h(z,t)dtdz+\left[h(z,t) - f(x,y,z)\right]dz^2+ dx^2+dy^2
\end{equation}
}
where $h(z, t)$ and $f(x,y,z)$ are arbitrary functions. The coordinates $(x, y, t)$ take all real values, while $z$ is taken to be periodic, $0\leq z \leq L$, with the endpoints identified such that $z=0$ and $z=L$ represent the same point.

The metric functions are fixed by the vacuum Einstein equations $G_{\mu \nu}=0$. The conditions $G_{xx}=G_{yy}=0$ determine $h(z,t)$ to be a traveling wave $h(z,t)\propto \exp[k(\epsilon z-\alpha t)]$, where $\epsilon$ and $\alpha$ are constants and $k$ may be real or imaginary, giving an exponential or an oscillatory profile respectively(however, the oscillatory case may fail to define a Lorentzian metric)~\cite{notesolu}. The equation $G_{zz}=0$ then reduces to the two-dimensional Laplace equation for $f(x,y,z)$, implying that $f(x,y,z)$ is a harmonic function in the transverse $(x,y)$ plane. The corresponding Riemann tensor has the non-vanishing components $2R_{izjz}=f(x,y,z)_{,ij}$ where $i$,$j~ \in \{x,y\}$, while all other components vanish. Thus, for a generic harmonic function $f(x,y,z)$, the spacetime is curved. In the special case $f(x,y,z)=0$, the Riemann tensor vanishes identically, and the metric becomes locally flat and isometric to a vacuum plane-fronted wave spacetime~\cite{2003-Stephani-book}. Note that the metric is everywhere regular with $det(g)=-h^{2}(z,t)$, provided $h(z,t)\neq0$. Therefore, no local geometric pathology is associated with the choice $f(x,y,z)=0$.

One immediately observes that the occurrence of CTCs in the metric (\ref{ansatz}) depends on the sign of $g_{zz}$. For fixed $x$, $y$, and $t$, this component is $g_{zz}=h(z,t) - f(x,y,z)$. When $h(z,t)>f(x,y,z)$ one has $g_{zz}>0$, and the closed curves of constant $x$, $y$, and $t$,  the $z-$ circles are spacelike. Conversely, when $h(z,t) < f(x,y,z)$, $g_{zz}<0$, and these closed curves become timelike, thereby forming CTCs. Thus, for a given harmonic function $f(x,y,z)$, the causal character of the closed $z-$ circles are governed entirely by the function  $h(z,t)$. We shall employ this vacuum solution as the core geometry of a new class of time-machine models satisfying the requirements outlined above. The spacetime evolution begins on a regular spacelike hypersurface (a partial Cauchy surface) that is asymptotically flat, topologically trivial, and satisfies the weak, dominant, and strong energy conditions. As  $h(z,t)$ evolves, a transition from $h(z,t)>f(x,y,z)$ to $h(z,t)<f(x,y,z)$ can occur, causing the closed $z$-circles to become timelike, with the extent of the resulting region depending on the tuning of $f(x,y,z)$ and $h(z,t)$. One may therefore envisage that an advanced civilization could manipulate the spacetime geometry so as to drive the transition  $h(z,t)>f(x,y,z)\rightarrow h(z,t)<f(x,y,z)$, thereby inducing the formation of CTCs.

%%%%%%%%%%%%%%%%% Flux %%%%%%%%%%%%%%%%%%%%%%%

%
\begin{figure}[h]
    \centering
    \includegraphics[width=0.42\textwidth]{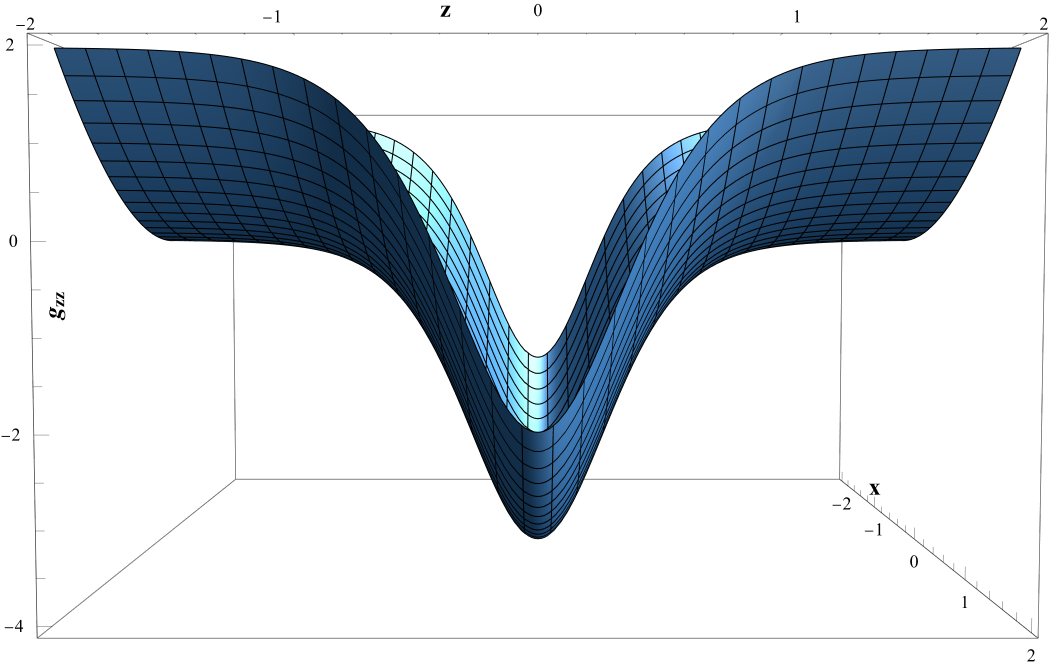}
    \caption{Model for the formation of CTCs based on Eq.~\eqref{model}.}
    \label{fig:CTCs}
\end{figure}
Fig.~\eqref{fig:CTCs} illustrates one such choice, for which we take
{\small
\begin{align}\label{model}
  h(z,t)=\operatorname{sech}^2\!\big(k( z- 2 t)\big),\qquad
  f(x,y,z)=\frac{x^2-y^2}{2},
\end{align}
}

with  $t=0$, and $k=2$. Varying $t$ merely shifts the location at which $g_{zz}$ vanishes in the $(x,y)$-plane, whereas varying $k$ sharpens or broadens the region where $g_{zz}<0$. 
The $g_{zz}<0$ region can therefore serve as the core of the time
machine~\cite{2005-Ori-PRL}. In this construction the causality
violation is confined to the vacuum core, where $G_{\mu\nu}=0$ and no
matter is present. The wave $h(z,t)$ that drives the transition must be
supported by a source lying outside this core, which we do not fix here;
we do not identify it with any established fundamental field, and its
evolution and the matching to an external asymptotically flat region lie
outside the scope of this work. What the vacuum solution establishes
is geometric: the empty core alone carries the onset of causality
violation.

%We emphasize, however, that this model is not unique: different choices of the function $h(z,t)$ may yield equally valid constructions. The matter content sourcing this geometry — which in turn constrains the form of $h(z,t)$ — resides in the region surrounding the vacuum core and may require a form of matter not yet associated with any specific known field. Nevertheless, the onset of causality violation arises within the internal vacuum core and is fixed by the initial conditions.

%%%%%%%%%%%%%%%%%%%%%%%%%%%%%%%%%%%%%%%%%%%%%%%%%%%

We require the CTCs to develop from a compact, regular spacelike
hypersurface $S_0$ in the vacuum core, rather than from a singularity or
from infinity. They cannot lie within the future Cauchy development $D^{+}(S_0)$, since
every point of $D^{+}(S_0)$ is causally determined by $S_0$ and admits
no closed causal curve. The most that can occur is a
closed null geodesic on its boundary, the future Cauchy horizon
$H^{+}(S_0)$~\cite{1973-Hawking_Ellis-Book}. This corresponds to the optimal causal
relation between $S_0$ and the causality violation for non-compactly
generated models, established by Ori~\cite{2005-Ori-PRL}. In our model, choosing $S_0$ with $h(z,t)>f(x,y,z)$ everywhere, so that the
$z$-circles are spacelike and the data are regular,
asymptotically flat, and respect the weak, dominant, and strong energy
conditions, the closed null geodesic $\gamma$ emerges as a generator of
$H^{+}(S_0)$ once the traveling wave drives the axis to $g_{zz}=0$, and
any regular extension across $\gamma$ drives $h(z,t)<f(x,y,z)$, rendering the $z$-circles
timelike. Because $H^{+}(S_0)$ is a Cauchy horizon of non-compact data and not a
compactly generated chronology horizon, Hawking's
theorem~\cite{1992-Hawking-PRD} does not apply; what $S_0$ determines is
the onset of causality violation, not the CTC region beyond the horizon.
%%%%%%%%%%%%%%%%% Flux %%%%%%%%%%%%%%%%%%%%%%%

The chronology horizon of Eq.~(\ref{ansatz}) is generated by a closed
null geodesic. Holding $x=y=t=0$ and letting $z$ traverse its period,
the orbit $\gamma$ of $\partial_z$ is closed, and it is null precisely
where the axis meets the horizon $g_{zz}=0$; it is a geodesic provided
the axis is a (necessarily saddle) critical point of the harmonic
function $f(x,y,z)$. Going once around $\gamma$ produces a boost,
\begin{equation}
  h_{\rm boost}=\frac{k(2\epsilon-\alpha)L}{2},
\end{equation}
so that for $\alpha\neq2\epsilon$ the geodesic is future incomplete, 
infinitely many windings within finite affine length, while remaining
in the compact circle and encountering no curvature singularity~\cite{1992-Hawking-PRD}. 

At $\alpha=2\epsilon$ the
boost vanishes and the single on-axis generator $\gamma$ of $H^{+}(S_0)$ closes up smoothly into a complete
closed null geodesic. The optical scalars of the horizon generators vanish identically,
$\rho=\sigma=\theta=0$, a direct consequence of the rigid, $z$-independent
transverse metric $dx^2+dy^2$: the generators neither expand, shear, nor
rotate. The per-circuit area factor and distortion therefore vanish and
the associated energy-condition inequality is saturated, consistent with
$R_{ab}=0$. Combining this with the vanishing boost, the degenerate tuning
$\alpha=2\epsilon$ realizes $\rho=\sigma=\theta=h_{\rm boost}=0$ on $\gamma$, which are exactly the conditions for the horizon tangent to
admit a Killing spinor. This is the unique circumstance in which Hawking's
vacuum-polarization divergence need not enforce chronology
protection~\cite{1992-Hawking-PRD}. These conditions are necessary but not sufficient: a finite $\langle T_{ab}\rangle$ would impose additional requirements on the quantum state, and since $f_{\rm area}=h_{\rm boost}=0$ the classical stability bound $f_{\rm area}>2h_{\rm boost}$ is only marginally satisfied(See appendix A).
% ------------------------------------------------------------------
% ------------------------------------------------------------------

\section{Conclusions} 
%%%%%%%%%%% Summary and Discussion %%%%%%%%%%%%%%%%%%

We conclude that, in our model, the onset of causality violation, the appearance of the closed null geodesics, is dictated entirely by the initial conditions specified on the compact vacuum region of the initial hypersurface, through the profile of the vacuum traveling wave. In this respect our construction is consistent with the analyses of Tipler and Ori, in which the closed null geodesics generating the chronology horizon are future-incomplete even though no local irregularity occurs along them~\cite{Tipler-1976-1977,1993-Ori-PRL}. We emphasize that the traveling-wave profile required for the transition from a spacelike to a timelike region, and its sustenance, may carry an energy cost that our purely classical vacuum analysis does not determine.

%We emphasize, however, that the traveling-wave profile does not by itself guarantee the formation of a closed-timelike-curve region: whether the transition from spacelike to timelike circles is actually realized, and sustaining it may carry an energy cost that our purely classical vacuum analysis does not determine.

Interestingly, we have identified at least one specific case in which
the single on-axis generator $\gamma$ of the chronology horizon becomes
a complete closed null geodesic of the type discussed by
Hawking~\cite{1992-Hawking-PRD,note}, while the supporting initial data
satisfy the weak, dominant, and strong energy conditions. This indicates
that the initial configuration required by our model is physically
acceptable, in the sense that it is in agreement with the classical
energy conditions. It is therefore conceivable that an advanced
civilization could engineer such initial data, and perhaps even that
natural processes involving the rapid propagation of these vacuum waves
through spacetime could approach the required configuration, though
such possibilities remain speculative.

The stability of time-machine solutions, including the model presented here, remains an open question, as many are unstable under classical perturbations or quantum fluctuations~\cite{1991-Kim_thone-PRD,1996-Kay_wald-CmMatPhy,1995-Krasnikov-1996}. Although these arguments were developed mainly for compactly generated models, some of the quantum-instability considerations extend to our construction as well, raising the question of whether a model like ours could be realized. These indications, however, do not rule out an actual construction: neither the strength of the quantum instability nor its eventual outcome, the spacetime configuration that finally forms, is currently known, and a definitive answer must await a full theory of quantum gravity.

%%%%%%%%%%%%%%%% Acknowledgements %%%%%%%%%%%%%
\begin{acknowledgements}
I thank Sumanta Chakraborty, Ashu Kushwaha, Avijit Chowdhury, Susobhan Mandal, Suvikranth Gera,  Tausif Parvez  for interesting and useful discussions. The research of SX is supported by Core Research Grants (CRG/2023/000934) from SERB.
\end{acknowledgements}
%%%%%%%%%%%%%%% Appendix %%%%%%%%%%%%%%%%
\appendix

\section{Conventions and the metric}\label{Append:01}
We use the signature $(-,+,+,+)$ and units $G=c=1$, and follow the
Newman--Penrose sign conventions of Ref.~\cite{1992-Hawking-PRD}. In
coordinates $x^{\mu}=(t,x,y,z)$ the line element is,
{\small
\begin{equation}\label{ansatza}
  ds^{2}=-2\,h(z,t)dtdz+\big[h(z,t)-f(x,y,z)\big]dz^{2}
         +dx^{2}+dy^{2},
\end{equation}
}
with $t,x,y\in\mathbb{R}$ and $z$ periodic, $0\le z\le L$, the
endpoints $z=0$ and $z=L$ being identified. The metric is Lorentzian wherever $h\neq0(\det g_{\mu\nu}=-h^{2}.)$; for the vacuum profile
obtained below, so~(\ref{ansatza}) is regular and
nondegenerate throughout. Two consequences are used repeatedly: $g_{tt}=0$, so $\partial_{t}$ is
null everywhere; and $g^{zz}=0$, so the hypersurfaces $z=\mathrm{const}$
are null, with $\partial_{t}$ as their generator.

% ====================================================================
\subsection{The closed null geodesic }\label{sec:geo}
% ====================================================================

We now exhibit the closed null geodesic on which the chronology
horizon of~(\ref{ansatza}) is generated. Consider the curve $\gamma$
obtained by holding $t=x=y=0$ and letting $z$ traverse its period.
Since $z=0$ and $z=L$ are identified, $\gamma$ is closed, with tangent
$\xi^{\mu}=\delta^{\mu}_{\,z}$. Its causal character is fixed by the
single component
\begin{equation}
  g_{zz}\big|_{\gamma}=h(z,0)-f(0,0,z)=e^{k\epsilon z}-f(0,0,z),
\end{equation}
so that $\gamma$ is null if and only if
\begin{equation}\label{eq:nulla}
  f(0,0,z)=e^{k\epsilon z}=h(z,0),
\end{equation}
i.e.\ provided the axis $x=y=0$ lies on the horizon $g_{zz}=0$ at
$t=0$. Because harmonicity constrains $f$ only in the
transverse plane, its profile along the $z$ axis is otherwise free and
condition~(\ref{eq:nulla}) may always be imposed.

To show that $\gamma$ is moreover a geodesic, note that for a curve
whose only nonvanishing velocity component is $\dot{z}$ the geodesic
equation collapses to
$\ddot{x}^{\mu}+\Gamma^{\mu}_{\,zz}\dot{z}^{2}=0$.
The transverse components vanish along $\gamma$ provided
\begin{equation}\label{eq:crit}
  \partial_{x}f=\partial_{y}f=0
  \qquad\text{on}\qquad x=y=0,
\end{equation}
i.e. the condition $\partial_{x}f=\partial_{y}f=0$, implies that the axis $x=y=0$ is a critical point of the transverse harmonic function $f$. Since $f$ satisfies  $\partial^2_{x}f+\partial^2_{y}f=0$, the transverse Hessian
\begin{align}
    H_{ij}=\partial_{i}\partial_{j}f \quad i,j\in\{x,y\} 
\end{align}
is trace-free. Consequently its eigenvalues sum to zero and, for a non-degenerate critical point, must have opposite signs. The critical point therefore cannot be a local maximum or minimum and is necessarily a saddle. The component $\Gamma^{t}_{\,zz}$ vanishes automatically: the null condition~(\ref{eq:nulla}) annihilates the term proportional to
$(f-h)$, while differentiating~(\ref{eq:nulla}) along the axis yields
$\partial_{z}f=\partial_{z}h$ there, cancelling the remainder. Thus,
under~(\ref{eq:nulla}) and~(\ref{eq:crit}), the closed curve $\gamma$ is
a null geodesic.

The remaining coefficient does not deflect $\gamma$ from the slice but
fixes its parametrization. For the vacuum profile Eq.\eqref{ansatza}, it is constant,
\begin{equation}
  \Gamma^{z}_{\,zz}=\frac{k(2\epsilon-\alpha)}{2}\equiv\kappa ,
\end{equation}
so $z$ is not affine. Integrating $\ddot{z}+\kappa\dot{z}^{2}=0$ gives
$dz/d\lambda\propto e^{-\kappa z}$, whence $\lambda\propto e^{\kappa z}$:
a single circuit $z\to z+L$ rescales the affinely normalized tangent by
$e^{\kappa L}$. This is the analogue of the boost of
Ref.~\cite{1992-Hawking-PRD},
\begin{equation}\label{eq:boosta}
  h_{\rm boost}=\kappa L=\frac{k(2\epsilon-\alpha)L}{2}.
\end{equation}
Whenever $\alpha\neq2\epsilon$ the boost is nonzero and $\gamma$ is
affinely incomplete in one direction, infinitely many windings
within finite affine length, while remaining inside the compact
circle and encountering no curvature singularity; with the time
orientation for which $h_{\rm boost}>0$ this is the future direction.
The geodesic $\gamma$ is therefore future incomplete in the sense of
Hawking~\cite{1992-Hawking-PRD}. In the
degenerate case $\alpha=2\epsilon$ the boost vanishes, $z$ becomes
affine, and $\gamma$ is complete.

% ====================================================================
\subsection{Null tetrad and optical scalars}\label{sec:tetrad}
% ====================================================================

To compute the convergence and shear of the horizon generators we
introduce, on the surface $g_{zz}=0$, the null tetrad
$(l^{a},n^{a},m^{a},\bar m^{a})$ adapted to $\gamma$,
\begin{align}
  l^{a}&=(\partial_{z})^{a}, &
  n^{a}&=h^{-1}(\partial_{t})^{a}, \\[2pt]
  m^{a}&=\tfrac{1}{\sqrt2}(\partial_{x}+i\,\partial_{y})^{a}, &
  \bar m^{a}&=\tfrac{1}{\sqrt2}(\partial_{x}-i\,\partial_{y})^{a},
\end{align}
with dual one-forms
\begin{equation}
  l_{a}dx^{a}=-h\,dt+(h-f)\,dz,\quad
  n_{a}dx^{a}=-dz,\quad
  m_{a}dx^{a}=\frac{1}{\sqrt2}\big(dx+i\,dy\big).
\end{equation}
On the horizon, where $h=f$, the first reduces to $l_{a}dx^{a}=-h\,dt$.
The tetrad obeys
\begin{equation}
  l^{a}l_{a}=n^{a}n_{a}=m^{a}m_{a}=0,\quad
  l^{a}n_{a}=-1,\quad m^{a}\bar m_{a}=1,
\end{equation}
all remaining contractions vanishing because~(\ref{ansatza}) contains no
$dt\,dx$, $dt\,dy$, $dz\,dx$, or $dz\,dy$ terms. The generator
$l^{a}=(\partial_{z})^{a}$ is null on $g_{zz}=0$ and geodesic, and is
affinely parametrized precisely at $\alpha=2\epsilon$; the cross-null
vector $n^{a}=h^{-1}(\partial_{t})^{a}$ is null throughout the
spacetime, since $g_{tt}=0$.

With the conventions of Ref.~\cite{1992-Hawking-PRD} the convergence and
shear are
\begin{equation}
  \rho=-\,m^{a}l_{a;c}\,\bar m^{c},\qquad
  \sigma=-\,m^{a}l_{a;c}\,m^{c},
\end{equation}
and the rotation $\theta$ of the dyad $\{m,\bar m\}$ along $l$ follows
from $\bar m^{a}m_{a;c}\,l^{c}$. For the transverse indices that enter
these contractions, $a,c\in\{x,y\}$, the covariant derivative of $l_{a}$
reduces to
\begin{equation}
  \nabla_{c}l_{a}=h\,\Gamma^{t}_{\,ca}-(h-f)\,\Gamma^{z}_{\,ca},
\end{equation}
and both $\Gamma^{t}_{\,ca}$ and $\Gamma^{z}_{\,ca}$ vanish identically,
since they involve only the (absent) mixed metric components and the
(constant) transverse block. Hence, for arbitrary $h(z,t)$ and harmonic
$f$,
\begin{equation}\label{eq:optical}
  \rho=0,\qquad \sigma=0,\qquad \theta=0 .
\end{equation}
These vanish identically, not merely on $\gamma$: the transverse
two-metric $dx^{2}+dy^{2}$ is rigid and independent of $z$, so the
congruence of generators neither expands, shears, nor rotates in the
screen space spanned by $m$ and $\bar m$. Equivalently the per-circuit
area factor and distortion of Ref.~\cite{1992-Hawking-PRD},
\begin{equation}
  f_{\rm area}=-2\oint\rho\,dt=0,\qquad q=-2\oint\sigma\,dt=0,
\end{equation}
both vanish, so the inequality
$\oint R_{ab}l^{a}l^{b}\,dt\le
-\big[h_{\rm boost}f_{\rm area}+\tfrac12(f_{\rm area}^{2}+q\bar q)\big]$
of Ref.~\cite{1992-Hawking-PRD} is saturated, in accord with the vacuum
condition $R_{ab}=0$.

Combining~(\ref{eq:optical}) with the boost~(\ref{eq:boosta}), at the
degenerate tuning $\alpha=2\epsilon$ the closed null geodesic $\gamma$
satisfies
\begin{equation}\label{eq:killing}
  \rho=\sigma=\theta=h_{\rm boost}=0
\end{equation}
simultaneously. These are exactly the conditions under which the
tangent to the horizon corresponds to a Killing spinor~\cite{1992-Hawking-PRD}.

%%%%%%%%%%%%%%%%%%%%%%%%%%%%%%%%%%%%
%%%%%%%%%%%%%%%%%%%%%%%%%%%%%%%%%%%%

%%%%%%%%%%%%%%%%%%%%%%%%%%%%%%%%%%%%

%uncomment the following line for two column reference.
%\twocolumngrid
%%%%%%%%%%%%%%% References %%%%%%%%%%%%%%%%
%\bibliographystyle{unsrt}
\bibliography{Reference} 

\begin{thebibliography}{18}
\bibitem [{()}]{notesolu}%
  \BibitemOpen
  \href@noop {} {}\bibinfo {note} {The general solution for $h(z,t)$ is $h(z,t)=P(z)\,Q(z-2t)$, where $P$ and $Q$ are arbitrary functions of their respective arguments.}\BibitemShut {Stop}%
\bibitem [{not()}]{note}%
  \BibitemOpen
  \href@noop {} {}\bibinfo {note} {Even at the degenerate tuning $\alpha=2\epsilon$, where the single on-axis generator $\gamma$ is a complete closed null geodesic, $H^{+}(S_0)$ remains non-compactly generated. Compact generation constrains the full generating congruence, not a single orbit: the off-axis generators rule the surface $g_{zz}=0$, i.e.\ $f(x,y,z)=h(z,t)$. Since $f$ is harmonic in the transverse $(x,y)$ plane, at each $z$ a non-constant $f$ has unbounded level curves, so the generators reach the non-compact transverse infinity and no compact set contains them all.}\BibitemShut {Stop}%
\end{thebibliography}
%\input{References.bib}
%\input{References.bbl}
%%%%%%%%%%%%%%%%%%%%%%%%%%%%%%%%%%%%%%%%%%
\end{document}